# Symmetry changes at the ferroelectric transition in the multiferroic YMnO$_3$

Gwilherm Nenert[1], Yang Ren[2], Harold T. Stokes[3], and Thomas T. M. Palstra[1]

[1] Solid State Chemistry Laboratory, Materials Science Centre, University of Groningen,
Nijenborg 4, 9747 AG Groningen, The Netherlands
[2] Experimental Facilities Division, Advanced Photon Source,
Argonne National Laboratory, Argonne, Illinois 60439, USA and
[3] Department of Physics and Astronomy, Brigham Young University, Provo, Utah 84602-4650, USA

We have identified, for the first time, the change in symmetry at the ferroelectric transition $T_{FE}$ near 1023K of the ferroelectromagnet YMnO$_3$. This transition takes place 300K below the transition to the centrosymmetric state at $T_{IP}$. Single crystal synchrotron diffraction coupled to a group theoretical analysis show that the paraelectric intermediate phase between $T_{IP}$ and $T_{FE}$ has P6$_3$/mcm symmetry. This proves that YMnO$_3$ is a proper ferroelectric and not an improper ferroelectric, as suggested by a previous group theoretical assignment. The origin of the ferroelectricity is caused by a correlated tilting of the MnO$_5$ polyhedra along the (100), (110) and (010) directions.

Coexistence of ferroelectricity and magnetism presents a rapidly growing and fascinating field in solid state physics. The hexagonal manganites RMnO$_3$ exhibit such coexistence with $T_{FE}$ ~ 1000K and $T_N$ ~ 100K. They show a ferroelectric transition between a high-temperature (HT) hexagonal centrosymmetric structure and room temperature (RT). Recent studies of YMnO$_3$ show that the ferroelectric and antiferromagnetic order parameters are coupled [1, 2]. In 2002 Fiebig et al. observed the coupling between the ferroelectric and magnetic domains [3]. Moreover, Lorentz et al. showed the existence of a novel reentrant phase and a ferroelectric-magnetic coupling in HoMnO$_3$ [4]. This allows the manipulation of electric and magnetic moments by magnetic and electric fields, respectively. Further interest in YMnO$_3$ has been generated, because it is considered as a non-volatile memory device element, and it can be grown epitaxially on SiO$_2$ [5]. Therefore, it has become of significant interest to understand the microscopic mechanisms of the ordering of all hexagonal RMnO$_3$.

While the interplay between electric and magnetic order in the low temperature phase becomes more clear, the origin of the ferroelectric state is under active investigation [6-11]. In several publications the origin of the ferroelectricity is suggested to be generated by a temperature dependent tilt of the MnO$_5$ bipyramids and a buckling of the R-layers. This is associated with a transition from the HT hexagonal centrosymmetric structure to the polar hexagonal symmetry at RT, in agreement with early conjectures by Lukaszemicz et al. [6]. However, measurements of the pyroelectric current [7] of YMnO$_3$ showed that the $T_{FE}$ ~ *933K*, which is several hundred degrees lower than the transition to the HT centrosymmetric state at $T_{IP}$ ~ 1273 K, observed by X-ray diffraction [6]. Therefore, Ismailzade et al. proposed the existence of an Intermediate Phase (IP), between the aforementioned structures. However, diffraction experiments up to 1073 K did not observe a phase transition to the IP [8]. In this Letter we provide experimental evidence for the existence of this IP using single crystal synchrotron-based diffraction data. Furthermore, we will show experimentally and group theoretically that the symmetry of the IP is different from previous analyses [9-11]. As a result we can distinguish between the complex polar displacements at $T_{FE}$ and the non polar ones at $T_{IP}$. The atomic displacements at $T_{FE}$ when considered apart, indicate a different nature of the ferroelectric transition than when considering the combined displacements resulting from the transitions at both $T_{IP}$ and $T_{FE}$.

YMnO$_3$ powder was synthesized by reacting stoichiometric amounts of Y$_2$O$_3$ (4N) and MnO$_2$ (3N metal basis) in nitrogen atmosphere at 1200°C for 10h. The powder was reground and resintered once under the same conditions to improve crystallinity. A single crystal was grown from the powder in air by the Floating Zone technique using a four mirror furnace. The crystallininity of the crystal was checked by Laue diffraction. The high quality of the crystal is further evidenced by narrow diffraction peaks ($\Delta 2\theta$ ~ 0.012$^0$ at RT). The diffraction experiments were performed on beamline BESSRC-11-ID-C at the Advanced Photon Source using a wavelength of 10.772 pm. The sample was heated in air using a four mirror furnace. We note that experiments by other groups were performed in vacuum, which may result in off-stoichiometry or even partial decomposition of the sample. The data were analyzed using the FIT2D program [12] to convert our image plate data to a 2D powder pattern, and then by the GSAS software package [13]. For comparison, the cell parameters have been refined in the P6$_3$cm setting at all temperatures.

In Figure 1, we show the temperature dependence of the lattice parameters. The lattice parameter c shows two phase transitions, one at about 1020 K and the other at about 1273 K. We associate the former with the ferroelectric transition and the latter with the transition to the centrosymmetric state. The symmetry changes are identified by temperature dependent single crystal diffraction. The inset of Figure 1 shows four diffraction patterns from RT up to 1353 K. The circles indicate the vanished Bragg peaks of the HT centrosymmetric phase. At 1283K, we can still observe some diffuse scattering. At 1353K, we can clearly observe the vanishing of Bragg intensity at the transition to the centrosymmetric state, resulting from the tripling of the unit cell.

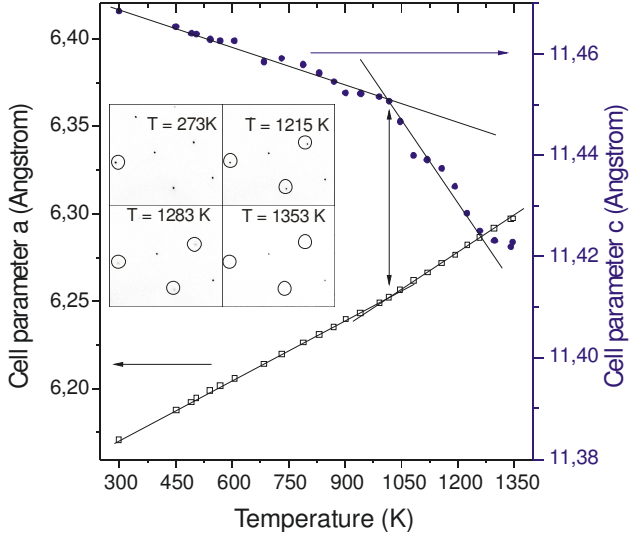

FIG. 1: Evolution of the cell parameters *a* (left) and c (right) with temperature. The solid lines are a guide for the eyes. The arrows indicate the ferroelectric transition temperature $T_{FE}$ and the tripling of the unit cell $T_{IP}$. The inset presents image plate data, showing the evolution of the peak intensities versus temperature around $T_{IP}$.

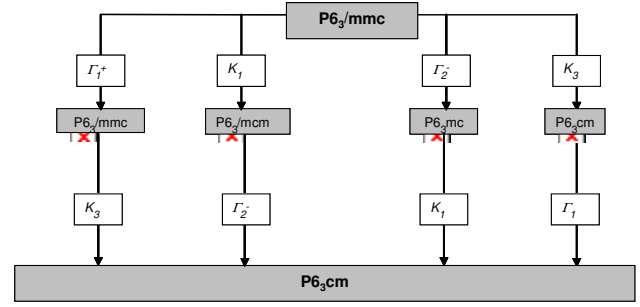

FIG. 2: Description of all group theoretically allowed phase transitions from $P6_3/mmc$ to $P6_3cm$ with an one intermediate phase. Only the primary order parameters are displayed. The associated wave vectors **k** are $\Gamma_i$ (**k** = 0, 0, 0) and $K_i$ (**k**= 1/3,1/3,0).

Recently, the crystal structure of $YMnO_3$ was studied in detail [8, 9, 14]. Van Aken et al. showed that the $Mn^{3+}$ are located in the centre of the coordinating O-bipyramid. However, their analysis neglected the observation that the ferroelectric transition takes place at considerably lower temperature than the transition into the centrosymmetric state. Therefore, this analysis is incomplete. Lonkai et al. measured the high temperature phase of other $RMnO_3$, (R = Tm, Yb and Lu) but they failed to observe a structural change at the ferroelectric transition using neutron diffraction [10]. Nevertheless, they developed a self-consistent model of the mechanism for ferroelectricity in hexagonal $RMnO_3$. They associated the IP with the $P6_3cm$ symmetry, using a group theoretical analysis. We will show by a full group theoretical analysis involving also the secondary order parameters that the symmetry of the IP is $P6_3/mcm$, in agreement with our experimental results.

Lonkai et al. were correct to point out that two separate phase transitions are expected in the transition from the HT phase ($P6_3/mmc$) to the RT phase ($P6_3cm$) [10]. However, they only considered the primary order parameter on the basis of which they derived three possible pathways of the subsequent phase transitions. A full analysis should include coupling to all symmetry allowed secondary order parameters. Using the program ISOTROPY, we derive four possible pathways from the HT phase to the RT phase, with no symmetry restriction for the IP, as shown in Figure 2. In our description, we need to consider different irreducible representations (IR): the polar IR ($\Gamma_2^-$), two different IR's which both give rise to a tripling of the unit cell ($K_3$ and $K_1$) and one strain IR ($\Gamma_1^+$). Thus, we obtain four possible space-groups in the IP, namely the transitions driven by the IR pairs ($\Gamma_1^+$, $K_3$),($K_1$, $\Gamma_2^-$), ($\Gamma_2^-$, $K_1$) and ($K_3$,$\Gamma_1$)[15]. The wave vectors associated with the irreducible representations are: $\Gamma_i^j$ (k = 0,0,0) and $K_i$ (**k** = 1/3,1/3,0). The space-groups $P6_3/mmc$ and $P6_3mc$ as IP can be excluded because this pathway involves a tripling of the unit cell only at the ferroelectric transition, in disagreement with our experimental results. Moreover the third pathway can be excluded as it involves the polarization mode $\Gamma_2^-$ as the primary order parameter at the higher temperature transition. It is also not in agreement with the experimental results. Lonkai et al. excluded $P6_3/mcm$ as IP because this order parameter generates only tiny additional peaks to the diffractogram indexed by $P6_3/mmc$ and would be nearly impossible to detect at T>1000K. They conclude that the IP should have space-group $P6_3cm$. We think this argument is unconvincing and disagrees with group theory and experimental observations.

The space-group $P6_3cm$ does not contain an inversion center and in order to generate zero-polarization, as observed experimentally, Lonkai et al. assume the IP to be antiferroelectric. We investigated this possibility using an analysis similar to that of Stokes and Hatch for $BaAl_2O_4$ [16]. Four IR's are involved in the transition from $P6_3/mmc$ to $P6_3cm$ as $K_3$ subduces in $P6_3cm$ to $1\Gamma_1^+$, $1\Gamma_2^-$, $1K_1$ and $1K_3$. This means that the atomic displacements from the transition can be decomposed into four modes which have the symmetry of these IR's.

| HT | IP | RT |
|---|---|---|
| $P6_3/mmc$ | $P6_3/mcm$ | $P6_3cm$ |
| Y ($D_{3d}$) | $Y_1$ ($D_{3d}$) | $Y_1$ ($C_{3v}$) |
|  | $Y_2$ ($D_3$) | $Y_2$ ($C_3$) |
| Mn ($D_{3h}$) | Mn ($C_{2v}$) | Mn ($C_s$) |
| $O_1$ | $O_{11}$ | $O_{11}$ |
|  | $O_{12}$ | $O_{12}$ |
| $O_2$ | $O_2$ | $O_{21}$ |
|  |  | $O_{22}$ |

TABLE I: Correspondence of the different atoms in the different symmetries. In brackets, we indicate the local symmetry of the metal atoms.

In the paraelectric state, these modes are vibrational modes centered at the equilibrium positions. We associate the transition with the order parameters $\eta_1$, $\eta_2$, $\eta_3$, and $\eta_4$, corresponding with the time averaged displacement of the IR's $K_3$, $\Gamma_1^+$, $\Gamma_2^-$, $K_1$, respectively. The primary order parameter $\eta_1$, associated with the $K_3$ mode transforms $P6_3/mmc$ to $P6_3cm$, whereas the secondary order parameters $\eta_2$, $\eta_3$, and $\eta_4$ transform $P6_3/mmc$ to $P6_3/mmc$ (n°194), $P6_3mc$ (n°186) (without tripling of the cell) and $P6_3/mcm$ (n°193), respectively. Using symmetry considerations, we can find an expression for the free energy as a function of the order parameters [15]:

$$F = F_0 + F_{12}\eta_1^2 + F_{14}\eta_1^4 + F_{22}\eta_2^2 + F_{23}\eta_2^3 + F_{24}\eta_2^4 + F_{32}\eta_3^2 + F_{34}\eta_3^4 + F_{42}\eta_4^2 + F_{43}\eta_4^3 + F_{44}\eta_4^4 + A_{12}\eta_1^2\eta_2 + A_{13}\eta_1^2\eta_3 + A_{14}\eta_1^2\eta_4 \quad (1)$$

The expansion is limited to terms up to the fourth degree, and we include only coupling terms which are linear in the secondary order parameter and quadratic in the primary order parameter. At the transition, the coefficient $F_{12}$ becomes negative, and the minimum of F occurs for a nonzero value of $\eta_1$. However, the coupling terms are linear in $\eta_2$, $\eta_3$, and $\eta_4$ and their contribution to the free energy becomes nonzero as well. In particular, through the coupling term $A_{13}\eta_1^2\eta_3$ the polar IR $\Gamma_2^-$ becomes finite, and will generate a finite polarization. This result would contradict the experimental results of Ismailzade et al. Furthermore, the transition would take place within the same space-group, driven by the $\Gamma_1^+$ IR, and should thus be first order as observed for Pnma perovskites [17], for which we find no experimental support. Finally, in this model the FE would be driven by a secondary order parameter, and YMnO$_3$ would then be an improper ferroelectric. Improper FE exhibit a saturation polarization that is much smaller than for proper FE. BaAl$_2$O$_4$ ($P_s \sim 0.08$ μC/cm$^2$) [16], Mg$_3$B$_7$O$_{13}$Cl ($P_s \sim 0.08$ μC/cm$^2$) [18] and Tl$_2$Cd$_2$(SO$_4$)3 (Ps $\sim$ 0.07μC/cm$^2$) [19] have significantly smaller polarization than YMnO$_3$ ($P_s \sim 5.5$ μC/cm$^2$) [20].

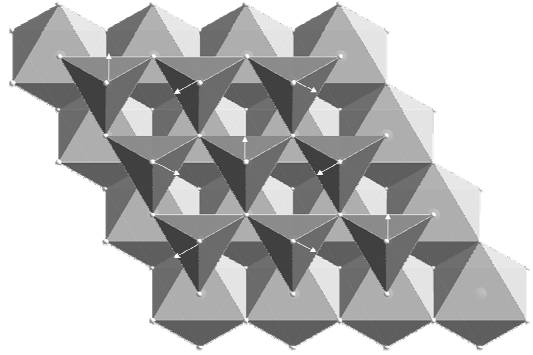

FIG. 3: Description of the atomic displacement of the tripling of the unit cell of YMnO$_3$ going from $P6_3/mmc$ to $P6_3/mcm$ at $T_{IP}$. The white arrows represent the displacement vectors of the oxygen O$_2$ (apical oxygens) and of the Mn atoms (not to scale).

We conclude that the space-group of the IP is $P6_3/mcm$, where the $K_1$ mode causes a tripling of the cell at $T_{IP}$, and the polarization mode $\Gamma_2^-$ yields to a proper ferroelectric state below $T_{FE}$. By deriving the Landau and Lifshitz criteria we predict that the transition should be first order at $T_{IP}$ and may be continuous at $T_{FE}$. While this is consistent with our data, the accuracy of the lattice parameters (Fig.1) is insufficient to prove this aspect. Moreover we can consider the secondary order parameters which are involved in both transitions ((1): transition $P6_3/mmc$ to $P6_3/mcm$ and (2): transition from $P6_3/mcm$ to $P6_3cm$). We find that the only secondary order parameters involved are strain modes ($\Gamma_1^+$ and $\Gamma_1$, respectively) but no polarization mode, because $K_1$ subduces in $P6_3/mcm$ $1K_1 + 1\Gamma_1^+$ and $\Gamma_2^-$ subduces in $P6_3cm$ $1\Gamma_2^- + 1\Gamma_1$.

Now that we have identified the correct symmetry of the different phases, we can describe the mechanism of the different transitions. We restrict our discussion to the displacement patterns associated with the primary order parameter only. The local symmetry for Y and Mn is given for the three phases in table 1. The displacements due to the $K_1$ mode from the HT phase to the IP are described in figure 3. We note that for this transition, mostly the apical oxygens O$_2$ and the Mn are involved. The other oxygen atoms O$_1$ are constrained by symmetry. In a trigonal pyramid, the O$_2$ and Mn atoms are displaced in the same direction. This displacement gives rise to an enhancement $\sqrt{3}a \times \sqrt{3}b$ of the unit cell. These distortions give rise to: 1) two inequivalent Mn -O distances, 2) remove the $C_3$ axis and the mirror plane containing the basal oxygen atoms, 3) lower the local environment of Mn from $D_{3h}$ to $C_{2v}$ These displacements create also two inequivalent Y - O distances and thus two inequivalent Y-sites.

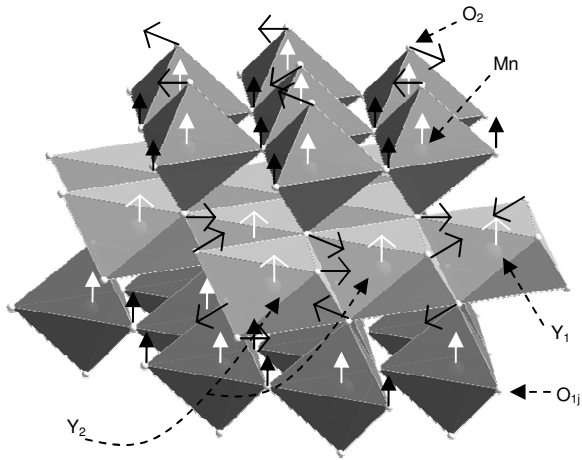

FIG. 4: Description of the atomic displacement pattern at the ferroelectric transition $T_{FE}$ of $YMnO_3$ going from $P6_3/mcm$ to $P6_3cm$. With black open arrows, we show the displacements of the apical oxygen $O_2$, with full black arrows the displacements of the basal oxygen $O_{1j}$, in white the displacements of the manganese (full arrows) and of the yttrium atoms (open arrows)(not to scale).

The displacements resulting from the ferroelectric mode $\Gamma_2^-$ below $T_{FE}$ are described in figure 4. The main feature of this transition is the creation of polarization along the $c$ axis. This polarization results from the correlated tilting of the $MnO_5$ polyhedra along the (100), (110) and (010) directions within the basal plane coupled to the displacement of the basal plane oxygen atoms along the $c$ axis. This tilting is accompanied by an in-plane displacement of the apical oxygens. This feature is noticeably different than in the scenario of van Aken et al. where the tilting results from an opposite displacement of the basal plane oxygen atoms of adjacent layers connected by the Y atom. The main contribution of the polarization is the displacement of one of the oxygen atoms of the basal plane of $MnO_5$ towards the coordination sphere of the yttrium atoms resulting in a signifcant di®erentiation in the Y - O(apical) distances. Thus, the coordination of the Y atoms changes from 6 to 6+1 . Also, the Y atoms within the $YO_6$ octahedra will come closer to one triangle of the oxygen octahedron and thus further away from the other triangle, resulting in a contribution to the polarization along the $c$ axis. Finally, the $C_2$ axis of the $MnO_5$ polyhedra is removed by the displacement of the Mn and the basal plane oxygens. These displacements are accompanied by a shortening of one $Mn-O_2$ distance while the other one is increased, thus, creating a polarization along the $c$ axis and two inequivalent $O_2$ oxygen atoms.

In conclusion, we observed for the first time using high resolution synchrotron data, two phase transitions in $YMnO_3$ associated with the ferroelectric transition and a tripling of the unit cell. Using a complete group theoretical analysis, we identify the transition from a centrosymmetric to a ferroelectric state for the hexagonal $RMnO_3$ as the succession of two paraelectric phases namely $P6_3/mmc$ and $P6_3/mcm$ and one ferroelectric phase with $P6_3cm$ symmetry. We prove that this family of compounds are not improper ferroelectrics but proper ferroelectrics in agreement with the magnitude of the polarization. The polarization can be described by a correlated zigzag tilting of the $MnO_5$ polyhedra due to the displacements of the apical oxygens instead of the basal plane oxygens.

ACKNOWLEDGEMENTS

We acknowledge stimulating discussions with M. Fiebig, T. Lonkai and D. Tomuta. This research is part of the MSC$^{plus}$ program. Use of the Advanced Photon Source was supported by the U. S. Department of Energy Office of Science, Office of Basic Energy Sciences, under contract No-W-37-109-Eg-38.